\documentclass[aps, prd, twocolumn, preprintnumbers, superscriptaddress, nofootinbib, floatfix, showpacs]{revtex4-1}

\usepackage{amsmath}
\usepackage{bm}
\usepackage{amsfonts}
\usepackage{graphicx}
\usepackage{epstopdf}
\usepackage{hyperref}
\usepackage{array}
\usepackage{soul}
\usepackage{color}
\usepackage{comment}
\enlargethispage{\baselineskip}

\hypersetup{
     colorlinks   = true,
     citecolor    = blue
}

\begin{document}

\title{\bf The QCD Axion and Gravitational Waves in light of NANOGrav results}

\author{Nicklas Ramberg}\email[Electronic address: ]{nramberg@uni-mainz.de}
\affiliation{PRISMA+ Cluster of Excellence \& Mainz Institute for Theoretical Physics, Johannes Gutenberg-Universit{\"a}t Mainz, 55099 Mainz, Germany}
\author{Luca Visinelli}\email[Electronic address: ]{luca.visinelli@lnf.infn.it}
\affiliation{INFN, Laboratori Nazionali di Frascati, C.P. 13, 100044 Frascati, Italy}
\date{\today}

\begin{abstract}
The North American Nanohertz Observatory for Gravitational Waves (NANOGrav) collaboration has recently reported strong evidence for a stochastic process affecting the 12.5 yr dataset of pulsar timing residuals. We show that the signal can be interpreted in terms of a stochastic gravitational wave background emitted from a network of axionic strings in the early Universe. The spontaneous breaking of the Peccei-Quinn symmetry originate the axionic string network and the QCD axion, the dark matter particle in the model. We explore a non-standard cosmological model driven by an exotic scalar field $\phi$ which evolves under the influence of a self-interacting potential; the axion field starts to oscillate during the modified cosmology, and provides the dark matter observed. For an equation of state $w_\phi < 1/3$, the QCD axion mass is smaller than expected in the standard cosmology and the GW spectrum from axionic strings is larger. We assess the parameter space of the model which is consistent with the NANOGrav-$12.5\,$yr detection, which can be explained within 95\% limit by a QCD axion field evolving in a dust-like scenario, as well as within 68\% limit in a cosmology with $w_\phi < 0$.
\end{abstract}
\preprint{MITP-20-077}
\maketitle

\section{Introduction}

The successful discover of gravitational waves (GWs) by the LIGO/Virgo collaborations of black holes and neutron stars coalescence have boosted the search for signals in various GW frequency ranges, which will become accessible within the next decade. A key aim of the upcoming GW searches consists in the detection of a stochastic GW background (SGWB) at frequencies $f\sim (1-100)\,$nHz, corresponding to the sensitivity of pulsar timing array (PTA) experiments such as the European Pulsar Timing Array~\cite{Lentati:2015qwp}, the Parkes Pulsar Timing Array~\cite{Shannon:2015ect} and the North American Nanohertz Observatory for Gravitational Waves (NANOGrav) collaboration~\cite{McLaughlin:2013ira, Brazier:2019mmu}. PTA data allow probing the SGWB in the nHz frequency range by keeping track of the correlated GW fluctuations from millisecond pulsars at the time of arrival. The presence of a SGWB in the nHz frequency window is generally expected from a variety of models~\cite{Caprini:2018mtu, Christensen:2018iqi} including supermassive black hole mergers~\cite{Sesana:2004sp}, phase transitions in the early Universe~\cite{Caprini:2010xv}, primordial magnetic fields~\cite{Pandey:2019tmo, Pandey:2020gjy}, and cosmic strings~\cite{Allen:1990tv, Siemens:2006yp, BlancoPillado:2011dq, Blanco-Pillado:2013qja, Blanco-Pillado:2017oxo, Chang:2019mza}.

Recently, the NANOGrav collaboration reported the evidence for a stochastic process from analyzing 12.5 years based on 45 pulsars~\cite{Arzoumanian:2020vkk}. The reported signal may be interpreted as a SGWB signal of amplitude $A_* \sim \mathcal{O}(10^{-10}-10^{-15})$ for frequencies $f \sim 30\,$nHz and a mild spectral index. At present, it is not clear whether the detected signal truly originates from a SGWB process due to i) the tension with previous PTA SGWB upper limits in the same frequency range, and ii) the lack of quadrupole correlations, a smoking gun for SGWB~\cite{Hellings:1983fr}.

If the detected signal found by the NANOGrav collaboration is indeed a component of the SGWB, it could be explained by various processes such as PBH formation~\cite{Vaskonen:2020lbd, DeLuca:2020agl}, phase transitions in the early Universe~\cite{Nakai:2020oit, Ratzinger:2020koh}, models of inflation in the early Universe~\cite{Vagnozzi:2020gtf, Domenech:2020ers}, or the GW signal emitted by a network of cosmic strings~\cite{Ellis:2020ena, Blasi:2020mfx, Samanta:2020cdk}. In particular, cosmic strings emit over a vast range of GWs, making these models particularly appealing for searches since fingerprints such as the spectral tilt or the strain amplitude can be probed across various GW windows in the near future~\cite{Sousa:2016ggw, Auclair:2019wcv}.

It is intriguing to explore models that are both able to predict the SGWB while at the same time providing a dark matter (DM) candidate, a missing tile in the cosmic puzzle. For example, the introduction of a new global symmetry in the early Universe which undergoes a spontaneous breaking could generate both a string network and the light Goldstone bosons that could act as the DM. The most notable case occurs for the QCD axion predicted within the theory of Peccei and Quinn (PQ)~\cite{Peccei:1977ur, Peccei:1977hh}, which could be the component of the observed DM in the Universe~\cite{Weinberg:1977ma, Wilczek:1977pj, Abbott:1982af, Dine:1982ah, Preskill:1982cy}.\footnote{See Refs.~\cite{Raffelt:1995ym, Raffelt:2006rj, Sikivie:2006ni, Kim:2008hd, Wantz:2009it, Kawasaki:2013ae, Marsh:2015xka, Kim:2017yqo, Irastorza:2018dyq, Sikivie:2020zpn, DiLuzio:2020wdo} for reviews of the QCD axion.} The spontaneous breaking of the global chiral symmetry that originates the axion field is accompanied by a network of strings that lasts as long as the axion is massless~\cite{Vilenkin:1981kz, Vilenkin:1984ib, Vilenkin:1986ku, Davis:1986xc, Davis:1989nj}. During its evolution, the axionic string network organizes into closed string loops which vibrate and release a spectrum of axions that also contribute to the DM budget~\cite{Vilenkin:1982ks, Harari:1987ht, Hagmann:1990mj, Chang:1998tb, Vilenkin:2000jqa}, as confirmed by precise cosmological simulations that attempt to assess the value of the axion mass~\cite{Klaer:2017qhr, Klaer:2017ond, Gorghetto:2018myk, Vaquero:2018tib, Buschmann:2019icd, Gorghetto:2020qws}.\footnote{Earlier work on the evolution of axionic strings is in Refs.~\cite{Battye:1993jv, Battye:1994qa, Battye:1994au, Martins:1995tg, Yamaguchi:1998gx, Yamaguchi:1999dy, Hiramatsu:2010yn, Hiramatsu:2010yu, Hiramatsu:2012sc, Hiramatsu:2012gg}.}

The value of the DM axion mass is altered if the axion field begins to oscillate when the Universe was not dominated by radiation in its early stages~\cite{Visinelli:2009kt}. Before Big-Bang nucleosynthesis (BBN) occurred, the Universe could have been filled by a yet undetected massive particle or some other form of exotic component, whose energy density could have driven a non-standard cosmology (NSC) period. For a fixed value of the axion mass $m_a$, the axion field evolving in an NSC would have a different energy density and velocity distribution than what expected in the standard cosmological scenario~\cite{Visinelli:2017imh, Visinelli:2018wza, Visinelli:2018utg, Draper:2018tmh, Nelson:2018via, Ramberg:2019dgi, Blinov:2019rhb}.

The NSC period would also modify the evolution of a global string network produced from a spontaneous symmetry breaking in the early Universe~\cite{Cui:2018rwi, Cui:2019kkd, Gouttenoire:2019kij, Blasi:2020wpy}. Primordial GW emission from cosmic string offers a probe to explore the content of the Universe at epochs before BBN, since the GW spectral tilt would depends on the detail of the NSC. For instance, the imprint of an early matter-dominated NSC could partially explain the result reported by NANOGrav through the phase transition that occurs at reheating~\cite{Bhattacharya:2020lhc} or through the different slope predicted in the power spectrum~\cite{Hook:2020phx}.

Although in the standard cosmological picture, the strain of the GW emission from axionic strings is generally too small to be detected even with the next generation of experiments, the spectrum of relic GWs is potentially amplified if an early NSC affected the evolution of the axion field evolution. In Ref.~\cite{Ramberg:2019dgi}, we have assessed the abundance of axion DM and the potential SGWB signature resulting from the axionic string network experiencing an NSC, focusing on the detection forecasts. If the equation of state describing the NSC is milder than the one of a relativistic component, a copious contribution to GWs might lead to a potentially detectable SGWB signal that could hint at the existence of axions jointly with direct detection of axion DM. Our treatment for the evolution of the string network thus differs in two key aspects from the previous literature: i) the axionic string network decays when the axion field acquires a mass and does not persist until the present time; ii) in our model, GW emission is a subdominant mechanism of energy release with respect to axion emission.

In this work, we provide a possible explanation of the NANOGrav detection in light of the model we studied in Ref.~\cite{Ramberg:2019dgi}, where the QCD axion is the CDM and GW emission from axionic strings are studied in light of an NSC in the early Universe. Our work intends to shed light on the possible models of the QCD axion and of the NSC that has the capability of being probed in future GW detectors. 

The paper is organized as follows. We introduce non-standard cosmological models in Sec.~\ref{sec:review}, and we review the axion model in Sec.~\ref{sec:QCDaxion}. The method to compute the GW spectrum within the theory is explained in Sec.~\ref{sec:method}. We show the results in Sec.~\ref{sec:nanograv_detection}, which are further discussed in Sec.~\ref{sec:discussion}. Conclusions are drawn in Sec.~\ref{sec:conclusions}.

\section{Setup of the cosmological model}
\label{sec:review}

In the standard picture, the expansion of the Universe is governed by the energy density of the relativistic bath produced right after inflation, until radiation is red-shifted away and matter domination begins. Within this picture, BBN successfully reproduces the abundance of various light elements with extreme precision, provided that the standard cosmological model holds up to temperatures $T_{\rm BBN} \sim 5{\rm \, MeV}$~\cite{Kawasaki:1999na, Kawasaki:2000en, Hannestad:2004px, ichikawa:2005vw, DeBernardis:2008zz, Gerbino:2016sgw}. The content of the Universe for temperatures $T > T_{\rm BBN}$ has not been explored, since a relic from the pre-BBN period has yet to be identified. One such relic could be the DM particle if it decouples from the plasma, or the GW spectrum released by some process in the pre-BBN era. Proposed probes comprise the effects on the chemical~\cite{Gelmini:2006mr, Gelmini:2008sh, Erickcek:2011us, Waldstein:2016blt, Visinelli:2017qga} and kinetic~\cite{Visinelli:2015eka} decoupling temperatures of the weakly interacting massive particle, as well as the altered energy density of thermal~\cite{Grin:2007yg} and non-thermal axions~\cite{Visinelli:2009kt}. Gravitational waves from early phase transitions or from topological defects could also be a gateway to explore the pre-BBN epoch~\cite{Giovannini:1998bp, Riazuelo:2000fc}.

Here, we model the pre-BBN era as follows. Soon after inflation ends, the expansion rate of the Universe is dominated by an exotic (non-radiation) component $\phi$, whose energy density $\rho_\phi$ is larger than that of the relativistic species at temperature $T$. We refer to this early stage as the NSC period. Candidates for the exotic component which is responsible for the NSC period include massive moduli fields~\cite{Dine:1982ah, Steinhardt:1983ia, Turner:1983he, Scherrer:1984fd} and fast ``kination'' fields~\cite{Barrow:1982ei, Ford:1986sy, Spokoiny:1993kt, Joyce:1996cp, Salati:2002md, Profumo:2003hq}.

The NSC period lasts until the exotic component either dilutes or decays away. Here, we focus on this latter case in which the details of the NSC are determined by the value of the decay rate $\Gamma$ and the equation of state $w_\phi$ of the exotic component. We treat the equation of state as a free parameter ranging over $-1/3 < w_\phi < 1/3$, excluding a post-inflation accelerated epoch where $w_\phi <-1/3$. We also do not consider the case in which the exotic fluid redshifts faster than radiation $w_\phi > 1/3$, because in this scenario the energy density of axionic strings is not enhanced with respect to the standard results~\cite{Ramberg:2019dgi}.

For a massive scalar field, the shape of the self-interacting potential determines the value of $w_\phi$. For example, the equation of state for a massive moduli field moving in the potential $V(\phi) \propto \phi^{2j}$ with $j>0$ is $w_\phi = (j-1)/(j+1)$~\cite{Turner:1983he}, so that for $j=1$ the field rolls in a quadratic potential and the dust-like case $w_\phi = 0$ is recovered. The effective equation of state for a massive field can attain negative values when the massive field dominates the expansion rate under the exponential potential~\cite{Wetterich:1987fm, Burd:1988ss, Copeland:1997et, Ferreira:1997hj}
\begin{equation}
    \label{eq:stringpotential}
    V(\phi) = V_0 \,\exp\left(-\lambda\phi/M_{\rm Pl}\right)\,,
\end{equation}
where $V_0$ and $\lambda$ are constant and $M_{\rm Pl}$ is the reduced Planck mass. The self-interaction potential of the form as in Eq.~\eqref{eq:stringpotential} arises in string models for the moduli fields associated with the geometry of the extra dimensions~\citep[e.g.\ Ref.][]{Green:1987sp}, in theories of modified gravity~\cite{Whitt:1984pd, Barrow:1988xh, Wands:1993uu}, or from supersymmetry breaking in models of gaugino condensation~\cite{Derendinger:1985kk, Dine:1985rz, deCarlos:1992kox}. In particular, for $2 < \lambda^2 < 3$, the Universe expands with the equation of state $-1/3 < w_\phi < 0$~\cite{Ratra:1987rm, Copeland:1997et}.

Depending on the nature of the exotic component, its interaction with radiation can be described by the Lagrangian term $\mathcal{L} \propto g\phi\bar\psi\psi$ if $\phi$ is a massive scalar field, where $\psi$ is the spinor describing an electron and $g$ a new coupling. Here, we do not include the details of the coupling between radiation and the exotic component $\phi$; instead, we describe the conversion of the energy density $\rho_\phi$ into radiation $\rho_R$ through energy conservation as~\cite{Giudice:2000dp, Giudice:2000ex, Giudice:2001ep, Visinelli:2014qla, Visinelli:2016rhn, Freese:2017ace}
\begin{eqnarray}
    \dot\rho_\phi &=& - 3(1+w_\phi)H\rho_\phi -\Gamma\rho_\phi\,, \label{eq:Kinetic_psi}\\
    \dot\rho_R &=& - 4H\rho_R + \Gamma\rho_\phi\,, \label{eq:Kinetic_Rad} \\
    3H^2 &=& 8\pi G(\rho_\phi + \rho_R)\,. \label{eq:Hubble}
\end{eqnarray}
The decay rate $\Gamma$ regulates the conversion rate of the exotic component into radiation. This set of equations describes a NSC in which the energy density $\rho_\phi$ dominates the expansion rate of the pre-BBN Universe before decaying into radiation, which thermalizes on timescales $\ll 1/\Gamma$. When $\rho_\phi$ equates the energy density in radiation, the Universe transitions to the radiation-dominated period at the temperature $T_\phi$. In order not to alter the results of BBN, we require $T_\phi > T_{\rm BBN}$. 

\section{The QCD axion}
\label{sec:QCDaxion}

The QCD axion is a hypothetical pseudo-scalar particle of zero-temperature mass~\cite{Weinberg:1977ma}
\begin{equation}
	\label{eq:axionmass}
	m_a = 6.2 {\rm \,\mu eV}\left( \frac{10^{12}{\rm GeV}}{f_a/N_{\rm DW}}\right)\,,
\end{equation}
where $f_a$ is the axion decay constant and $N_{\rm DW}$ is the ``domain wall number'', see e.g. Sec.~2.7.1 in Ref.~\cite{DiLuzio:2020wdo}. Here, we set $N_{\rm DW} = 1$. The mass of the axion arises from QCD instanton effects and depends on temperature so that $m_a(T) = \sqrt{\chi(T)}/f_a$, where the QCD topological susceptibility is normalized at zero-temperature as $\chi(0) = m_a^2f_a^2$~\cite{Gross:1980br}.

The abundance of axions produced through non-thermal mechanisms after the PQ symmetry breaking occurs could explain the missing DM in the Universe~\cite{Preskill:1982cy, Abbott:1982af, Dine:1982ah}.\footnote{A thermal axion component is also expected from processes scattering off pions and nucleons~\cite{Berezhiani:1992rk}.} The computation of the present abundance proceeds through the vacuum realignment mechanism, for which the axion field $a$ in units of $f_a$, the so-called axion angle $\theta = a/f_a$, reads
\begin{equation}
	\label{eq:axioneqmotion}
	\ddot\theta + 3H\dot\theta + m_a^2(T)\sin\theta = 0\,.
\end{equation}
Given an initial value of the axion angle, $\theta_i$ drawn randomly from the uniform distribution $[-\pi, \pi]$ when the PQ phase transition occurs, the solution to Eq.~\eqref{eq:axioneqmotion} is a constant value of $\theta = \theta_i$ as long as the Hubble friction is much larger than the axion mass. Coherent oscillations in the axion field begin at around the time $t_{\rm osc}$ given by
\begin{equation}
	\label{eq:axioneqmotion1}
	H(t_{\rm osc}) \approx m_a(t_{\rm osc})\,,
\end{equation}
after which the number of axions in a comoving volume is fixed and the axion energy density evolves as a matter-like field. The temperature of the plasma at $t_{\rm osc}$ is $T_{\rm osc}$.

\subsection{Axions from strings} \label{sec:axion_strings}

If PQ symmetry broke either after inflation or was temporarily restored right after inflation, an emergence of topological defects that eventually decay will contribute to the DM axion budget~\cite{Davis:1985pt, Davis:1986xc}. The string network contains about one axionic string per Hubble volume and is approximated to have a linear mass distribution of string core size $\sim 1/f_a$ and a linear mass density~\cite{Vilenkin:1982ks, Davis:1986xc}
\begin{equation}
	\label{eq:mass_length}
	\mu_{\rm eff}(t) = \pi f_a^2\ln\left(f_a t\right)\,.
\end{equation}
The string network evolves by emitting a spectrum of axions and GW, either by wiggles on long open strings or self collapse of closed strings. The cold portion of the spectrum of axions emitted from axionic strings might significantly contribute to the present energy density of axions~\cite{Vilenkin:1981kz, Vilenkin:1982ks, Vilenkin:1984ib, Vilenkin:1986ku, Davis:1986xc, Harari:1987ht, Davis:1989nj, Battye:1993jv, Battye:1994qa, Battye:1994au, Martins:1995tg, Yamaguchi:1998gx, Yamaguchi:1999dy, Vilenkin:2000jqa, Hiramatsu:2010yn, Hiramatsu:2010yu, Hiramatsu:2012sc, Hiramatsu:2012gg, Klaer:2017qhr, Klaer:2017ond, Gorghetto:2018myk, Vaquero:2018tib, Buschmann:2019icd, Gorghetto:2020qws}.

To describe the power loss of the network into radiation, we consider the dissipation of the energy $E_{\rm loop} = \mu_{\rm eff} \ell$ of a closed loop with length $\ell$ into axions and gravitational waves~\cite{Battye:1993jv, Battye:1994qa, Battye:1994au, Martins:1995tg},
\begin{equation}
	\label{eq:decayrateloop}
	P_{\rm loop} = \frac{\mathrm{d}E_{\rm loop}}{\mathrm{d}t} = \kappa\mu_{\rm eff} + \gamma_{\rm GW}G\mu_{\rm eff}^2,
\end{equation}
where $\gamma_{\rm GW} \approx 65$ and $\kappa \approx \mathcal{O}(0.1)$ are \textit{dimensionless} quantities describing strings moving at relativistic speed~\cite{Vachaspati:1986cc, Sakellariadou:1990ne, Sakellariadou:1991sd}. Contrarily to the previous literature, we have set $\kappa \approx 0.15$ which characterizes the predominant energy loss into axions, instead of using the value $\kappa = 0$ which would describe the predominant release of energy into gravity wave modes. Since the ratio of the power loss in gravity waves and axions is of the order of $Gf_a^2 \ll 1$, the string network mainly dissipates energy into axions. While sub-Planckian, the value of $f_a$ is larger than in the standard scenario for an equation of state $w_\phi < 1/3$, leading to an enhanced GW emission.

Using Eq.~\eqref{eq:decayrateloop}, the shrinking of a loop with initial size $\ell_i$ is described by the expression
\begin{equation}
	\label{eq:shrinking}
	\frac{\mathrm{d}\ell}{\mathrm{d}t} = \kappa - \ell\,\frac{\mathrm{d}\ln\mu_{\rm eff}}{\mathrm{d}t},
\end{equation}
where $\ell = \ell(t, \ell_i)$ is the size at time $t$. Although the loop length could vary between arbitrary sizes, numerical simulations show that the initial length of the large loop at its formation tracks the time of formation as $\ell(t_i) =\alpha t_i$, where $\alpha$ is an approximately constant loop size parameter which gives the fraction of the Hubble horizon size at which loops predominantly form~\cite{Battye:1994au, BlancoPillado:2011dq, Blanco-Pillado:2017oxo}.

Owing to the small power loss in GWs, we approximate the evolution of the string network in a scaling regime with the emission proceeding through axions. The energy density of the radiated axions follows the evolution $\rho_a + 4H\rho_a = \Gamma_{\rm str \to a}$, where $\Gamma_{\rm str \to a}$ is the energy lost in the emission of axions per unit time. The number density of axions emitted from strings within the modes of angular wavenumber $k \approx 1/\ell(t_i) \approx H(t_i)/\alpha$ to infinity is~\cite{Gorghetto:2018myk}
\begin{equation}
	\label{eq:numberdensityaxions}
	n_a^{\rm str} = \int^t\,\mathrm{d}t'\frac{\Gamma_{\rm str \to a}(t')}{H(t')}\left(\frac{R(t')}{R(t)}\right)^3\,\int\frac{\mathrm{d}k}{k}\,F(k),
\end{equation}
where $R(t)$ is the scale factor at $t$ and the spectral energy density is defined in terms of a spectral index $q>1$ as~\cite{Battye:1993jv, Battye:1994qa, Gorghetto:2018myk}
\begin{equation}
	\label{eq_axionspectrum}
	F(k) = \frac{q-1}{\alpha^{q-1}}\left(\frac{k}{H}\right)^{-q}.
\end{equation}
The spectral energy density $F(k)$ is properly normalized over the frequency range considered.

It has been alternatively assumed that strings efficiently shrink emitting all of their energy at once, leading to a flat power spectrum per logarithmic interval with a harder spectral index $q=1$, an infrared cutoff at the wave mode $k \approx H$ and a ultraviolet cutoff at $k = f_a$~\cite{Harari:1987ht, Hagmann:1990mj, Chang:1998tb}.

\subsection{Gravitational Waves from Axionic String Loops}
\label{Gravitational Waves from Axionic String Loops}

Here, we compute the subdominant SGWB emitted from axionic string loops. The fraction of the critical energy density released into the GW spectrum per unit logarithmic interval of frequency is
\begin{equation}
	\Omega_{\rm GW}(t, f) = \frac{1}{\rho_c(t)}\frac{\mathrm{d}\rho_{\rm GW}}{\mathrm{d}\ln k}.
\end{equation}
The evolution of the string loop is described by Eq.~\eqref{eq:shrinking}, where the shrinking rate is not driven by the emission into GWs but rather into Goldstone bosons. GWs emitted at time $t'$ with the modal frequency $f_{\rm emit}$ redshift to $f = f_{\rm emit}R(t')/R(t)$ at a later time $t > t'$.

The axionic string network emits GWs as long as the axion is massless. When coherent axion oscillations begin, the string network dissipates due to the formation domain walls, and the energy density of GWs emitted so far redshifts as radiation to present time,
\begin{equation}
	\label{eq:OmegaGWred}
	\Omega_{\rm GW}(t_0, f_0) = \frac{\rho_c(t_{\rm osc})}{\rho_c(t_0)}\left(\!\frac{R(t_{\rm osc})}{R(t_0)}\right)^4 \Omega_{\rm GW}(t_{\rm osc}, f)\,,
\end{equation}
where $\rho_c(t)$ is the critical density of the Universe at time $t$ and the frequency $f_0 = f R(t_{\rm osc})/R(t_0)$ accounts for the redshift of the peak wavelength.

We decompose the fractional energy density of GWs emitted by the string network loops in terms of the distribution of power mode of emission $n$ as
\begin{equation}
	\label{eq:OmegaGW}
	\Omega_{\rm GW}(t_{\rm osc}, f) = \gamma_{\rm GW}\,\sum_n \frac{n^{-4/3}}{\mathcal{N}}\, \Omega_{\rm GW}^{(n)}(t_{\rm osc}, f)\,,
\end{equation}
where $\mathcal{N} = \sum_n n^{-4/3}$, the power spectrum of index $q=4/3$ characterizes the emission of GW modes from loops with cusps~\cite{Blanco-Pillado:2013qja, Blanco-Pillado:2017oxo}, and the contribution from the mode $n$ at time $t$ reads~\cite{Cui:2018rwi}
\begin{eqnarray}
	\label{eq:GWmode}
	\Omega_{\rm GW}^{(n)}(t_{\rm osc}, f) &=& \frac{1}{\rho_c(t)} \frac{2n}{f}\frac{\xi}{\alpha} \times \\
	&& \int_{t_s}^{t_{\rm osc}} \!\! \mathrm{d}t' \frac{G\mu_{\rm eff}^2(t')}{t_i^4}\left(\frac{R(t')}{R(t)}\right)^5\left(\frac{R(t_i)}{R(t')}\right)^3\,.\nonumber
\end{eqnarray}
In Eq.~\eqref{eq:GWmode}, the time $t_i$ at which the loop forms is obtained from inverting the redshift expression for the emitted frequency. Since the sum in the expression for the total emission converges slowly, higher emission modes significantly contribute to the total power. The thermal history of the Universe prior $t_{\rm osc}$, thus the choice of the NSC, enters Eq.~\eqref{eq:GWmode} through the scale factors appearing in the integrand.

\section{Method}
\label{sec:method}

We have solved numerically the set of coupled kinetic equations describing the decay of the exotic field $\phi$ into radiation in Eqs.~\eqref{eq:Kinetic_psi}-\eqref{eq:Hubble}, as a function of the parameters $(T_\phi, w_\phi)$ and assuming that the radiation energy density is negligible at temperatures well above $T_\phi$. We recover the time dependence of the Hubble rate $H$, which is fed into the equation of motion for the QCD axion field in Eq.~\eqref{eq:axioneqmotion} which describes the vacuum realignment mechanism (vrm). Since we are considering temperatures well above those at which matter-radiation equality occurs, the axion is a subdominant field during the pre-BBN epoch and does not appear in Eqs.~\eqref{eq:Kinetic_psi}-\eqref{eq:Hubble}.

The axion angle $\theta = a/f_a$, where $a$ is the axion field, evolves according to Eq.~\eqref{eq:axioneqmotion}, starting from the initial condition $\theta(t_i) = \theta_i$ and $\dot \theta(t_i) = 0$ (a dot is a derivation with respect to cosmic time) at some time $t_i$ well before the time $t_{\rm osc}$ at which the axion acquires a non-zero mass, see Eq.~\eqref{eq:axioneqmotion1}. The initial value of the axion angle is fixed as $\theta_i = \pi/\sqrt{3}$~\cite{Visinelli:2009zm, Visinelli:2014twa}. The evolution of the axion field is sensitive to the total energy content of the Universe through the value of $t_{\rm osc}$, which depends on the value of the Hubble rate $H$. If coherent oscillations begin during a NSC, the abundance of axions differs from the standard result for a given value of $f_a$. This is extremely relevant if the transition temperature $T_\phi$ lies below the GeV~\cite{Visinelli:2009kt}.

The present number of axions per comoving volume resulting from the vacuum realignment mechanism just described, $n_a^{\rm vrm}$, and from axionic string emission $n_a^{\rm str}$. For the string contribution, we have integrated Eq.~\eqref{eq:numberdensityaxions} numerically to include the time-varying linear mass density in Eq.~\eqref{eq:mass_length} that characterizes the axionic string. The total number density of cold axions gets contributions from both string decay and vacuum realignment mechanism, and the resulting present energy density of axions is a function of $f_a$, $w_\phi$, and $T_\phi$. The value of $f_a$ is then fixed by assuming that the axion is the DM particle, so that the value of the DM axion mass and the axion energy constant depend on the parameters $w_\phi$ and $T_{\rm RH}$~\cite{Ramberg:2019dgi}.

\section{The QCD axion and NANOGrav}
\label{sec:nanograv_detection}

The results from NANOGrav-$12.5\,$yr searches from PTA data~\cite{Arzoumanian:2020vkk} are reported in terms of the power-law spectrum of the characteristic strain
\begin{equation}
	\label{eq:strain}
	h_c(f) = A_* \left(\frac{f}{f_{\rm yr}}\right)^{\frac{3-\gamma}{2}}\,,
\end{equation}
where $f_{\rm yr} = 1{\rm \, yr^{-1}}$ is a reference frequency, $A_*$ is the amplitude at $f_{\rm yr}$, and the parameter $\gamma$ is related to the spectral tilt. The fractional energy density in GWs associated with the strain is~\cite{Ellis:2020ena}
\begin{equation}
	\label{eq:strain1}
	\Omega_{\rm GW}(f) = \frac{2 \pi^{2}}{3 H_0^2}f^{2}h_c^2(f) \equiv \Omega_{\rm GW}^{\rm yr} \left(\frac{f}{f_{\rm yr}}\right)^{5-\gamma},
\end{equation}
where $\Omega_{\rm GW}^{\rm yr} = 2 \pi^{2}f_{\rm yr}^{2}/(3 H_0^2)$.

The NANOGrav collaboration reports the fit of the strain in Eq.~\eqref{eq:strain} to thirty bins within the frequency range $f \in (2.5, 90)\,$nHz. However, the excess is reported by fitting only the first five bins in the signal-dominated frequency range $f \in (2.5, 12)\,$nHz, while bins of higher frequencies are assumed to constitute of white noise. The constraint derived on the $(A_*, \gamma)$ space of parameters reads approximately $\log_{10} A_* \in (-15.8, -15.0)$ and $\gamma \in (4.5, 6.5)$ at 68\% confidence level (CL).

We assess the cosmological scenario presented against these experimental results, considering the GW signal from axionic strings given in Eq.~\eqref{eq:OmegaGWred}. We obtain the spectral tilt and the amplitude by inverting the relations in Eqs.~\eqref{eq:strain}-\eqref{eq:strain1} as~\cite{Ellis:2020ena}
\begin{eqnarray}
	\gamma &=& 5 - \frac{\mathrm{d}\ln\Omega_{\rm GW}(t_0, f)}{\mathrm{d}\ln f}\bigg|_{f=f_*}\,,\\
	A_* &=& \sqrt{\frac{3H_0^2}{2\pi^2}\,\frac{\Omega_{\rm GW}(t_0, f_*)}{f_{\rm yr}^2}\,\left(\frac{f_{\rm yr}}{f_*}\right)^{5-\gamma}}\,,
\end{eqnarray}
where the quantities are computed at the reference frequency $f_* = 5.6\,$nHz which is the geometric average of the signal-dominated frequency range considered. 

\begin{figure}
	\includegraphics[width = 0.99\linewidth]{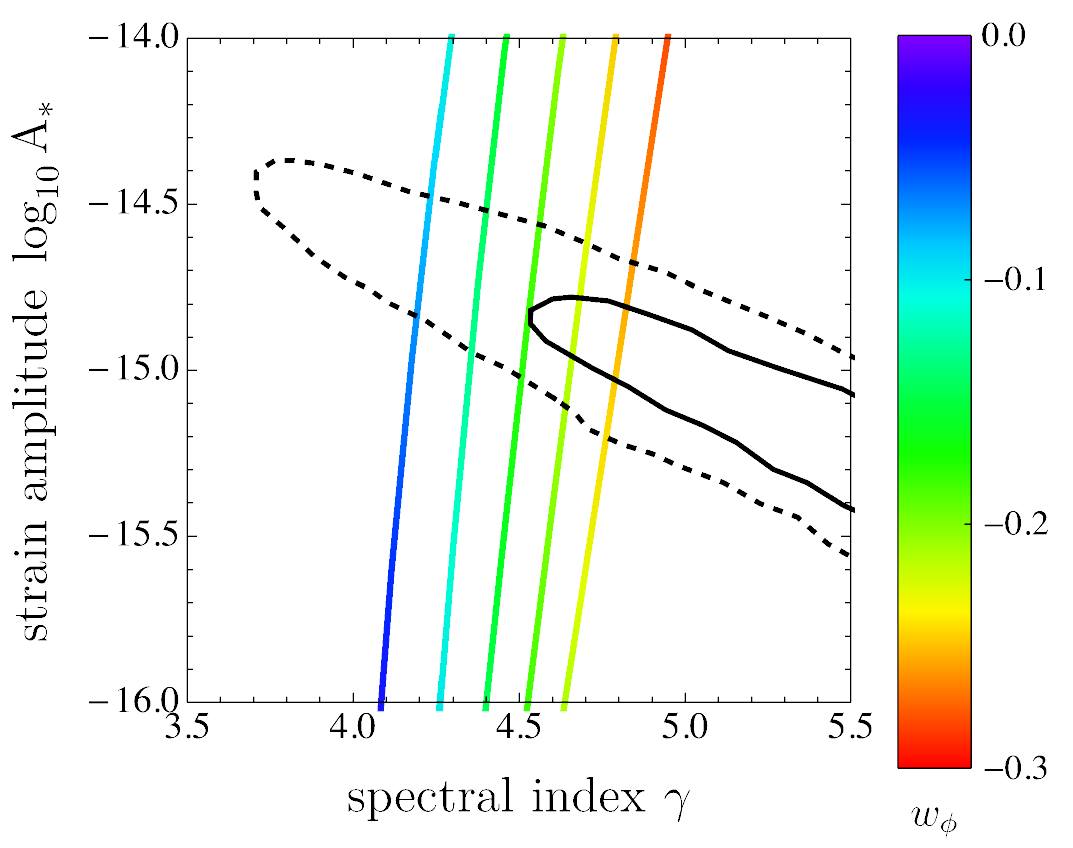} 
	\caption{The colored curves represent the amplitude $A_*$ (vertical axis) and the spectral tilt $\gamma$ (horizontal axis) of the SGWB predicted from axionic string, for different values of $T_\phi$. From left to right, $T_\phi \in (5, 10, 15, 20, 25)\,$MeV. The color codes different values of the equation of state $w_\phi$ along each line, as given by the color bar to the right of the figure. The solid and dashed black lines indicate the detection of ($A_*$, $\gamma$) respectively at 68\% and 95\%, as inferred by the analysis of the NANOGrav collaboration~\cite{Arzoumanian:2020vkk}.}
	\label{fig:wphi}
\end{figure}

In Fig.~\ref{fig:wphi} we compare the results of the NANOGrav collaboration with the prediction of the SGWB from axionic strings in our model. The constraints on the strain $A_*$ (vertical axis) and the spectral tilt $\gamma$ (horizontal axis) from the NANOGrav collaboration are shown at 68\% CL (solid black line) and 95\% CL (dashed black line). Each color curve is a prediction of the model for different values of $T_\phi$. From left to right, $T_\phi \in (5, 10, 15, 20, 25)\,$MeV. The color codes the value of the equation of state $w_\phi$ of the exotic component that controls the NSC, which varies within the region $w_\phi \in (-0.3, 0)$. Lines in blue represent models with a dust-like equation of state, $w_\phi \approx 0$, for which the model predicts a SGWB signal within 95\% limit reported with a relatively low value of the transition temperature $T_\phi \lesssim 10\,$MeV. For these models, the GW strain amplitude lies between $\log_{10} A_* \in (-14.9, -14.4)$ with a spectral index $\gamma \sim 4.3$. Lines with a red shade represent models with $w_\phi \lesssim -0.2$. For these models, the expected SGWB from the axionic string network is within 68\% limit, with the amplitude in the range $\log_{10} A_* \in (-15.3, -14.6)$ and a spectral tilt $\gamma \approx (4.7-4.8)$. The NANOGrav data then favor a negative equation of state $w_\phi \lesssim -0.2$.

Note, that the QCD axion is the DM particle in the model we consider, with the value of $m_a$ sensibly differing from the result expected in a standard cosmological model. In particular, for the target parameter space that reconciles the NANOGrav results, we expect $m_a \in (0.01, 1)\,\mu$eV~\cite{Ramberg:2019dgi}. In turns, the axion energy scale in our model is expected to be larger than the corresponding quantity in the standard cosmology, yielding to a sizable SGWB which could potentially be detected by next-generation detectors across different GW frequencies. The predictions of the model would need to be confirmed through a direct detection of the QCD axion, whose light mass is in reach of ``A Broadband/Resonant Approach to Cosmic Axion Detection with an Amplifying B-field Ring Apparatus'' (ABRACADABRA)~\cite{Kahn:2016aff, Ouellet:2018beu} and ``UPconversion Loop Oscillator Axion Detection experiment'' (UPLOAD)~\cite{2019arXiv190405774T, Thomson:2019aht}.

\section{Discussion}
\label{sec:discussion}

If the SGWB signal is confirmed, we could have a precious insight of the physics of the early Universe. More data is needed to confirm the result and, in case, to distinguish different models that candidate to explain the results in either astrophysical or cosmological setups. Models in which a spontaneous symmetry breaking leads to a string network whose GW emission can potentially explain the NANOGrav results in terms of a SGWB have been recently discussed in the literature~\cite{Ellis:2020gtq, Blasi:2020mfx}. Our model presents key differences from these models since i) the axionic string in our model does not last until present time and it is dissipated in the early Universe as soon as the QCD axion acquires a mass, and ii) the axionic string network predominantly emits axions before decaying, with a subdominant spectrum of GWs. While the GW spectrum from axionic strings can be generally neglected, the GW strain is potentially detectable in future detectors for the region of the parameter space we consider in this work, $w_\phi \leq 0$.

At present, no dedicated numerical simulation that accounts for the evolution of the axion field in a NSC exist. In particular, no simulations of the axionic string network in background cosmologies other than radiation ($w_\phi = 1/3$) and matter ($w_\phi = 0$) have been explored. For a mild equation of state of the exotic field $w_\phi < 1/3$, our results suggest that the emission of GWs, while negligible in the evolution of the string network, could lead to a complementary and detectable signal in near-future experiments, if the transition temperature lies at relatively low scales. This process might have been overlooked in the axion literature since the emission of GWs during the evolution of an axionic string can be safely neglected in the standard cosmological scenario, due to the strong suppression of the GW emission with respect to the dominant emission in axions.

Future exploration of the SGWB jointly with laboratory searches could shed light on the nature of the axion as the DM particle, as well as on the content of the Universe in its early stages. If the signal is confirmed and more data becomes available, a dedicated Monte Carlo analysis would pin down the preferred region in the parameter space of the model $(T_\phi, w_\phi)$, allowing us to gather deeper insights into the cosmological history of the early Universe. In turn, the analysis would ideally identify a preferred range in which the axion could be the DM particle which could be tackled by a dedicated laboratory search.

Further, an interest to bear in mind is to perform lattice simulations for axion strings in modified expansions histories of the early universe because of the unconventional results that GW's from axion strings are within detectable reach for a set of near-future experiments, even though their amplitudes are suppressed. Several conclusions emerge from this paper, firstly when more precise frequency binned data is present, evidence of a pre-BBN relic of the early universe might reveal itself and shed light on whether DM constitutes QCD axions. Secondly, the revelation of the axion theory can appear by considering its associated gravitational wave relics, which in fact might be the "smoking gun" because present axion experiments are very model-dependent.

\section{Conclusions}
\label{sec:conclusions}

In this paper, we have analyzed one class of cosmological models in which the QCD axion constitutes the DM particle. The Peccei-Quinn phase transition from which the axion originates takes place after inflation so that the associated topological defects are not washed away and constitute an axionic string network that evolves releasing energy into a spectrum of axions. If the evolution of the Universe around and below the GeV is non-standard, it is possible that a copious amount of GWs are also released from the string network, feeding into a stochastic background. We assessed the expected SGWB signal focusing on the frequency range $f \in (2.5, 12)\,$nHz where the signal dominates the pulsar timing, in light of the recent results from the NANOGrav-$12.5\,$yr data analysis~\cite{Arzoumanian:2020vkk}.

Our results are summarized in Fig.~\ref{fig:wphi}, where we show the prediction of the axionic string model proposed in light of the NANOGrav-$12.5\,$yr results (solid black line is for 68\% detection, and dashed black line is for 95\% detection), for different values of the transition temperature $T_\phi$ (colored lines) and of the equation of state for the exotic field governing the NSC (color scale). Our model predicts a SGWB within the 95\% limit of the PTA detection, when considering dust-like scenarios with $w_\phi \approx 0$ and a relatively low transition temperature $T_\phi \lesssim 10\,$MeV. For these models, the GW strain amplitude lies between $\log_{10} A_* \in (-14.9, -14.4)$ with a spectral index $\gamma \sim 4.3$. For the case of a background model whose equation of states satisfies $w_\phi \lesssim -0.2$, the expected SGWB from the axionic string network is within 68\% limit, with the amplitude in the range $\log_{10} A_* \in (-15.3, -14.6)$ and a spectral tilt $\gamma \approx (4.7-4.8)$. The preferred region of the parameter space hints at background models $w_\phi \lesssim -0.2$, with a transition temperature $T_\phi \lesssim 100\,$MeV. Although we have not specified the underlying model for the exotic component which drives the background cosmology during NSC, a theory for the self-interacting potential of a scalar field that leads to $w_\phi < 0$ has been presented in Eq.~\eqref{eq:stringpotential}.

Our result sheds light on the preferred mass regions of the DM axion below the $\mu$eV target, in range for the ABRACADABRA experimental setup~\cite{Kahn:2016aff, Ouellet:2018beu}. The detection of an axion of mass $\mathcal{O}(10^{-8})\,$eV could be complementary to the crossing evidences coming from probing the primordial GW wave spectrum. Detecting the QCD axion could come along with the possibility of detecting a SGWB and could provide a guideline for understanding the cosmology of the early Universe. This result contrasts the result for the QCD axion required to be the DM in the standard cosmology, where the axion mass is expected to be in the range $m_a \approx (10 - 500)\,\mu$eV, the uncertainty arising from the computations involving the decay of the axionic string network. A summary of the values for the DM axion mass in our model as a function of the parameters $(T_\phi, w_\phi)$ is found in Ref.~\cite{Ramberg:2019dgi}.

\begin{acknowledgments}
NR thanks Pedro Schwaller \& Wolfram Ratzinger for useful discussion and acknowledges support by the Cluster of Excellence ``Precision Physics, Fundamental Interactions, and Structure of Matter'' (PRISMA+ EXC 2118/1) funded by the German Research Foundation (DFG) within the German Excellence Strategy (Project ID 39083149). LV~acknowledges support from the NWO Physics Vrij Programme ``The Hidden Universe of Weakly Interacting Particles'' with project number 680.92.18.03 (NWO Vrije Programma), which is (partly) financed by the Dutch Research Council (NWO), as well as support from the European Union's Horizon 2020 research and innovation programme under the Marie Sk{\l}odowska-Curie grant agreement No.~754496 (H2020-MSCA-COFUND-2016 FELLINI). We thank Nordita, the Oskar Klein Centre for Cosmoparticle Physics at Stockholm University, and Uppsala University, where this line of research was started, for hospitality.
\end{acknowledgments}

\bibliography{NanoGRAV.bib}

\end{document}